\RequirePackage{fix-cm}
\documentclass[smallextended]{svjour3}       % onecolumn (second format)
\smartqed  % flush right qed marks, e.g. at end of proof
\usepackage{graphicx}
\usepackage{fullpage}
\usepackage{multirow}
 \newcommand{\amu}{a_{\mu}}
 
 \newcommand{\amuhvp}{\amu^{\mathrm{HVP}} }
 \newcommand{\GeV}{\mathrm{GeV}}

 \journalname{Few-Body Systems}
\begin{document}

\title{
~ {\hfill \small CERN-TH-2017-088}\\
Lattice calculations and the muon anomalous magnetic moment%\thanks{Grants or other notes
}


\author{Marina Krsti\'c Marinkovi\'c   
  %\and        Second Author %etc.
}


\institute{  Trinity College Dublin %\at
	    % 1211 Geneva 23, Switzerland
                     %  \\
           \hfill  Theoretical Physics Department \\
             School of Mathematics   \hfill CERN \\
	    Dublin 2, Ireland   \hfill  Geneva, Switzerland\\
 	    Tel.: +353-1-896-4598 \hfill Tel.:+41-2276-74222 \\
             Fax: +353-1-896-2282 \hfill Fax:  +41-2276-73850\\
              \email{mmarina@maths.tcd.ie} \hfill \email{marina.marinkovic@cern.ch}
}

\date{~% \hfill \bf CERN-TH-2017-088
}

\maketitle

\begin{abstract}
Anomalous magnetic moment of the muon, $a_{\mu}=(g_{\mu}-2)/2$, is one of the most precisely measured quantities in particle physics and it provides a stringent test of the Standard Model. The planned improvements of the experimental precision at Fermilab and at J-PARC propel further reduction of the theoretical uncertainty
of $\amu$.   
The hope is that the efforts on both sides will help resolve the current discrepancy between the experimental measurement of $a_{\mu}$ and its theoretical prediction, and potentially gain insight into new physics. 
The dominant sources of the uncertainty in the theoretical prediction of $a_{\mu}$ are the errors of the hadronic contributions. I will discuss recent progress on determination of hadronic contributions to $a_{\mu}$ from lattice calculations.
\keywords{lattice field theory \and muon anomalous magnetic moment \and hadronic vacuum polarization \and hadronic light by light}
 \PACS{11.15.Ha \and 12.38.Gc \and 13.40.Em}
\end{abstract}

\section{Introduction}
\label{sec:intro}
%Your text comes here. Separate text sections with
The contributions of new physics (NP) beyond the Standard Model (SM) to the anomalous magnetic moment of a lepton, $a_\ell=\frac{g_l-2}{2}$, are expected to scale
as $ a_\ell - a_\ell^{\rm SM} \propto (m^2_\ell / \Lambda^2_{\rm NP})$
for leptons $\ell=e,\mu,\tau$ and some new physics scale $\Lambda_{\rm NP}$. The muon magnetic moment, $a_{\mu}$, is
therefore
 $m_{\mu}^2/m_{e}^2$ times more sensitive to NP than the one of the electron, while the magnetic moment of $\tau$ is not yet experimentally accessible.  Additionally, $a_{\mu}$ can be  
obtained with high precision both from the experiment \cite{Bennett2006}, as well as from the SM prediction \cite{Jegerlehner2016,Davier2016} 
and as such  serves as one of the most sensitive tests of the SM.

The experimental measurement of 
$\amu$  
currently shows a $3 \sigma$  to $4 \sigma$ deviation from the 
SM estimate \cite{Jegerlehner2016,Davier2016}. 
The summary of various SM contributions to $\amu$  given in Table \ref{tab:amu} shows that the dominating uncertainty comes from the hadronic contributions. 
The value of the hadronic vacuum polarization (HVP) entering the theoretical average in Table \ref{tab:amu} is obtained via a dispersion integral of the total hadronic $e^+e^-$ annihilation and $\tau$-decays cross sections  (for comprehensive reviews see \cite{Passera2005,Jegerlehner2016}), while the value of hadronic Light by Light contribution (HLbL) quoted in Table \ref{tab:amu} is purely based on low energy effective QCD hadronic models. 
Due to the plans of Fermilab \cite{Venanzoni2016} and J-PARC \cite{Saito2012} experiments to reduce the experimental uncertainty by a factor of four, performing an independent check of the phenomenological estimate of the hadronic contributions is of crucial importance in order to
 confirm or resolve current discrepancy between theory and experiment, 
and potentially gain insight into new physics beyond the SM. 
Lattice discretization of gauge theories provides a way to compute the contributions to $\amu$ with prevailing uncertainties, HVP and HLbL,  from first principles. We will present in the following sections the ongoing progress of these calculations, aiming to achieve sub-percent precision for the HVP and 
$\approx 10\%$ precision for HLbL.
\begin{table}
\caption{Comparison between a SM prediction and the experimental result for $\amu$. The last column contains example Feynman diagrams for the individual contributions to the total theoretical estimate.
}
\label{tab:amu}       % Give a unique label
\centering
\begin{tabular}{llclclc|c}
\hline\noalign{\smallskip}
%\begin{small}
\vspace*{-0.5em} Contribution & Value  & Uncertainty & Sample Diagrams \\ 
%& diagrams  \\
\noalign{\smallskip}\hline\noalign{\smallskip}
\vspace*{-0.5em} QED $O(\alpha)$~~ & \hfill  116 140 973.21 & \hfill 0.03 & \vspace*{0em}\hspace*{-0.3em} \includegraphics[width=0.05\textwidth]{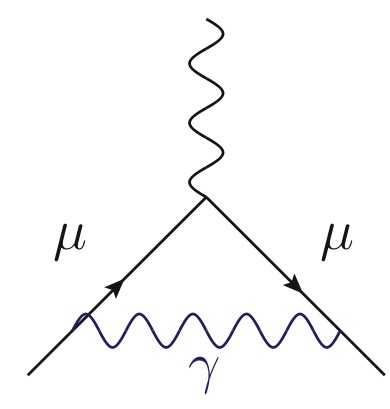}
\hspace*{0.5em} \\
QED $O(\alpha^2)$ &\hfill  413217.63 & \hfill 0.01& \includegraphics[width=0.05\textwidth]{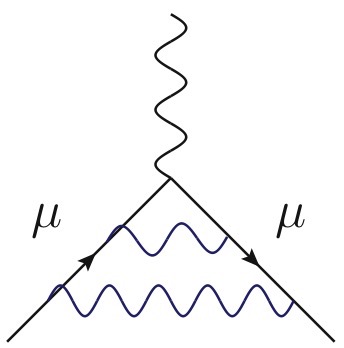}
\hspace*{0.5em} \includegraphics[width=0.05\textwidth]{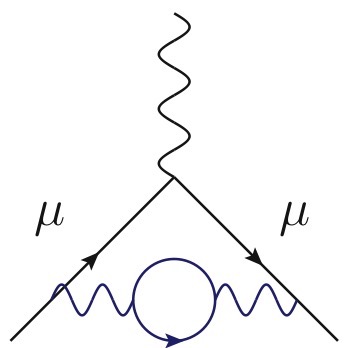} \\
QED $O(\alpha^3)$ &\hfill  30 141.90 &\hfill  0.00 &\hspace*{2.5em} \multirow{2}{*}{\vdots} \\
QED $O(\alpha^4)$ & \hfill  381.01& \hfill 0.02 &  \\[-1mm]
QED $O(\alpha^5)$ &\hfill 5.09 &\hfill  0.01 &\includegraphics[width=0.25\textwidth]{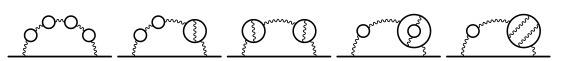} \\
%%%%%%%%%%%%%%%%%%%%%
QED, total \cite{Aoyama:2014sxa}& \hfill 116 584 718.95 & \hfill 0.04\\[-1mm]\noalign{\smallskip}\hline\\[-2mm]
 electroweak, total \cite{Gnendiger2013}& \hfill 153.6 &\hfill  1.0 &  \includegraphics[width=0.05\textwidth]{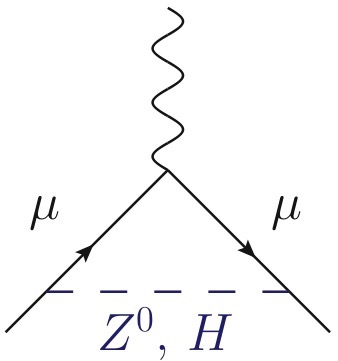}
\hspace*{0.5em} \includegraphics[width=0.05\textwidth]{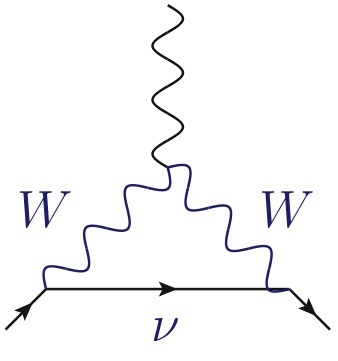}\\[-1mm]
\noalign{\smallskip}\hline\\[-3mm]
%%%%%%%%%%%%%%%%%%%%%
HVP (LO) \hfill \cite{Hagiwara:2011af} &\hfill 6949.9 &\hfill  43.0 & % \hspace*{2.5em} 
 \vspace*{0em}\hspace*{-0.3em} \includegraphics[width=0.05\textwidth]{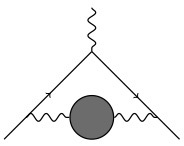}
\hspace*{0.5em} \\
%%%%%%%%%%%%%%%%%%%%%
HVP (NLO)  \hfill \cite{Hagiwara:2011af} &\hfill  -98.0 &\hfill  1.0  &  % \hspace*{2.5em}  
 \includegraphics[width=0.05\textwidth]{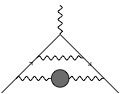}
\hspace*{0.5em} \includegraphics[width=0.05\textwidth]{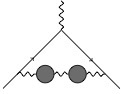} \\
%%%%%%%%%%%%%%%%%%%%%
HLbL \hfill  \cite{Jegerlehner2009} & \hfill 116.0 & \hfill 40.0 &  
 \vspace*{0em}\hspace*{-0.3em} \includegraphics[width=0.045\textwidth]{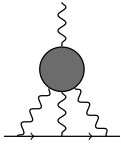} \\
%%%%%%%%%%%%%%%%%%%%%
HVP(NNLO)  \hfill  \cite{Kurz2014a} &\hfill  12.4 & \hfill  0.1 & %\hspace*{2.5em}
 \includegraphics[width=0.05\textwidth]{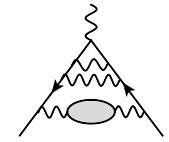}
\hspace*{0.5em} \includegraphics[width=0.05\textwidth]{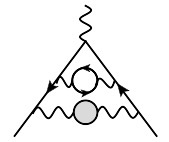} 
\hspace*{0.5em} \includegraphics[width=0.05\textwidth]{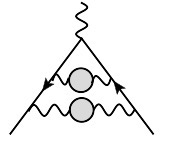} \\
%%%%%%%%%%%%%%%%%%%%%
HLbL(NLO) \hfill  \cite{Colangelo2014} & \hfill 3.0 &\hfill  2.0 & % \hspace*{2.5em}  
 \includegraphics[width=0.05\textwidth]{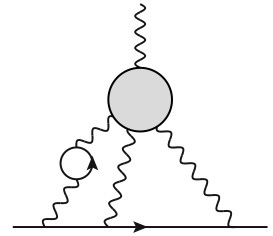} \\
\noalign{\smallskip}\hline 
theory, total & \hfill 116 591 855.0 & \hfill 59.0 & ~ \\
\noalign{\smallskip}\hline
experiment \cite{Mohr2016} & \hfill 116 592 089.0 &\hfill  63.0 & ~\\ 
\noalign{\smallskip}\hline
%\end{small}
\end{tabular}
\end{table} 

\section{Hadronic Vacuum polarization in Euclidean space-time} 
\label{sec:hvp}
\begin{figure}
\begin{minipage}{0.25\textwidth}
\centering
 \includegraphics[width=0.6\textwidth]{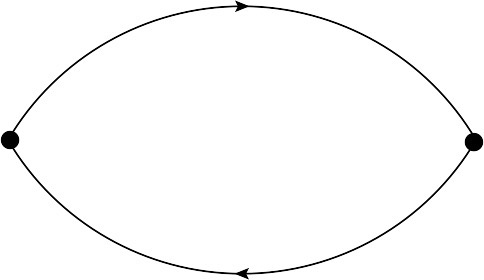}\\[3mm]
 \includegraphics[width=0.65\textwidth]{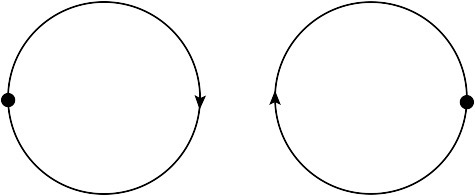}
\end{minipage}
\begin{minipage}{0.32\textwidth}
  \includegraphics[width=1.15\textwidth]{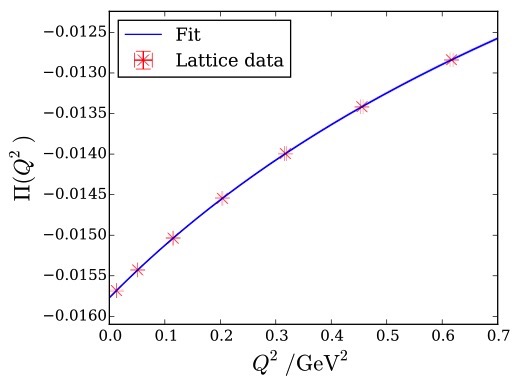}
\end{minipage}
\hspace*{0.05\textwidth}
\begin{minipage}{0.32\textwidth}
  \includegraphics[width=1.15\textwidth]{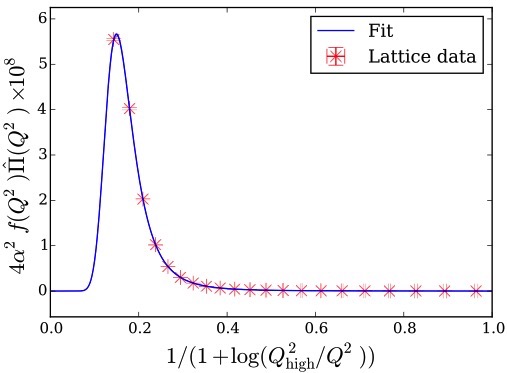}
\end{minipage}
\caption{Left panel: connected HVP (up) and disconnected HVP contribution (down). The dots represent the insertion of the conserved  and/or local vector currents. Central panel: strange HVP function ($\Pi_{\mu\nu}(Q)=a^4 \sum_{x} e^{iQx} \langle J_{\mu}^{em,s}(x) J_{\nu}^{em,s} (0) \rangle$) reproduced from Ref. \cite{Blum2016a}, measured on $N_f=2+1$ configurations with Domain Wall fermions. Right panel: the strange quark contribution to the integrand of Eq. \ref{eq:hvp}, reproduced from the strange HVP study of Ref. \cite{Blum2016a}.}
\label{fig:strange}       % Give a unique label
\end{figure}

Ever since the first attempt to perform a lattice QCD computation of the HVP is made by T. Blum \cite{Blum:2002ii}, a precise computation of this quantity has become one of the long-standing goals of the lattice comunity. The leading order HVP 
correction to $\amu$ at Euclidean space-time is obtained as \cite{Lautrup1968,Blum:2002ii}:
\begin{equation} 
\amuhvp =  \left(  \frac{\alpha}{\pi} \right)^2 \int_0^{\infty} dQ^2 f(Q^2) \times \hat{\Pi}(Q^2), 
\label{eq:hvp}
\end{equation}
where 
$\hat{\Pi}(Q^2)= {\Pi}(Q^2)- {\Pi}(0)$
is the %infra-red subtracted 
renormalized scalar vacuum polarization, $\alpha$ is the fine structure constant, $m_{\mu}$ is the muon mass
and $f(Q^2)$ is analytically known\footnote{$f(Q^2) = m_{\mu}^2 Q^2 Z^3(Q^2) \frac{1-Q^2Z(Q^2)}{1+m_{\mu}^2Q^2Z^2(Q^2)}$, where $Z(Q^2)=\frac{\sqrt{(Q^2)^2+4m_{\mu}^2Q^2}-Q^2}{2m_{\mu}^2 Q^2}$. } function. 
Gauge and Lorentz invariance relate scalar hadronic vacuum polarization function and vacuum polarization tensor  $\Pi_{\mu\nu}(Q)=(Q^2\delta_{\mu\nu}-Q_{\mu}Q_{\nu})\Pi(Q^2)$, which can be computed using a lattice discretization of the vector current $J^{em}_{\mu}$: 
\begin{equation}
\Pi_{\mu\nu}(Q)=a^4 \sum_{x} e^{iQx} \langle J_{\mu}^{em}(x) J_{\nu}^{em} (0) \rangle.
\label{eq:2pt}
\end{equation}
The vector two-point correlator in Eq. \ref{eq:2pt} consists of the quark-connected and quark-disconnected contributions illustrated in the leftmost panel of Fig. \ref{fig:strange}, whose continuum and infinite volume limits are usually computed independently. 

The integrand of Eq. \ref{eq:hvp} is strongly peaked at $Q^2 \approx \frac{m^2_{\mu}}{4}$ and the integral corresponding to the $\amuhvp$ is therefore dominated by the low momentum region (see  Fig. \ref{fig:strange}). Lattice computation is in practice performed in a finite box, with time and spatial extents $L_{\mu}$ and periodic boundary
conditions. The lattice momenta are thus quantized: $Q_{\mu}=\frac{2\pi}{L_{\mu}} n_{\mu}$, $n_{\mu}=0,\dots,L_{\mu}-1,\mu=0\dots3$, 
and this condition leaves us with few lattice points in the most important region for the evaluation of the HVP integral: around and bellow the peak of the integrand. The latter calls for the usage of models or parametrizations that would estimate the value of $\Pi(Q^2)$ in the low momentum region. 
Moreover, the signal to noise ratio worsens at small $Q^2$. 
For all these reasons, a number of tricks for noise reduction and control of the systematic uncertainties  needed to be developed in order to bring the accuracy of the lattice computation of $\amuhvp$ closer to the precision of the estimates coming from the phenomenology and experiment. 
%%%%%%%%%%%%%%%%
%%%%%%%%%%%%%%%%
%%%%%%%%%%%%%%%%
%%%%%%%%%%%%%%%%
\subsection{Strategies for the evaluation of the HVP on the lattice}
%%%%%%%%%%%%%%%%
%%%%%%%%%%%%%%%%
\label{sec:strategies}
%%%%%%%%%%%%%%%%
%%%%%%%%%%%%%%%%
Initially, computing $\amu^{HVP}$ in lattice QCD required
performing a fit of the lattice data in order to obtain the 
renormalized 
amplitude $\hat{\Pi}(Q^2)$ \cite{Blum:2002ii,Boyle2012,Aubin:2012me}.  
The fits were introducing model dependence that made systematics difficult to control. 
We discuss several recent proposals for the non-perturbative determinations of the HVP, 
whose development was motivated by overcoming the aforementioned difficulty of model-dependent extraction of the additive renormalization  ${\Pi}(0)$ and a slope of the HVP near $Q^2=0$.
%%%%%%%%%%%%%%%%
%%%%%%%%%%%%%%%%
%%%%%%%%%%%%%%%%
%%%%%%%%%%%%%%%%

\noindent
{\it  Pad\'e approximants.} 
%\paragraph{Pade approximants.}
In order to avoid uncontrolled systematics originating from performing model dependent fits of the HVP (usually based on Vector Meson Dominance),  
a model independent approach for constructing the functional form of $\Pi(Q^2)$ has been proposed by the authors of Ref. \cite{Aubin:2012me}. The suggested method is based only on known mathematical properties of the vacuum polarization, and consists of fitting a sequence of Pad\'e approximants to the lattice data. 
The Pad\'es are such that the full sequence converges to the actual polarization for any compact region in the complex plane excluding the cut along the negative real axis. 
Increasing the order of the Pad\'e approximation 
allows for testing the stability of the fits and guarantees a more reliable estimate of the systematics 
that come mainly from the decreasing precision of lattice data at low $Q^2$.\\

%%%%%%%%%%%%%%%%
%%%%%%%%%%%%%%%%
%%%%%%%%%%%%%%%%
%%%%%%%%%%%%%%%%
\noindent
{\it Time Moments.} 
%\paragraph{\it Time Moments} 
%%%%%%%%%%%%%%%%
%%%%%%%%%%%%%%%%
Recently proposed approach by the HPQCD collaboration  \cite{Chakraborty2014} obtains the HVP by reconstructing the Adler function from its derivatives. 
From the space-averaged current-current correlator 
$G(x_0)=-\frac{a^3}{3}\sum_{\vec{x}} \langle J_{\mu}^{em}(x) J_{\nu}^{em} (0) \rangle$, one  computes the time-moments of the renormalized HVP 
by taking the $Q^2$-derivatives at zero spatial momenta: $G_{2n}=(-1)^n \frac{\partial^{2n}}{\partial Q^{2n}} \{Q^2 \hat{\Pi}(Q^2)\}|_{Q^2=0}=a \sum_{x_0}x_0^{2n}G(x_0)$. 
The relation of the HVP form factor with the constructed time-moments allows us to determine the coefficients of a given HVP function  parametrization (Pad\'e approximants, conformal polynomials etc.). 
Further developments of this idea, which replace continuous time-moments with discrete- 
 and/or spatial-moments, have been applied in more recent works by  RBC/UKQCD, Mainz group and  BMW collaboration \cite{Blum2016a,HorchPoS,Lellouch2016}.\\
%Francis:2014hoa
%%%%%%%%%%%%%%%%
%%%%%%%%%%%%%%%%
%%%%%%%%%%%%%%%%
%%%%%%%%%%%%%%%%
%%%%%%%%%%%%%%%%
%%%%%%%%%%%%%%%%
%%%%%%%%%%%%%%%%

\noindent
{\it Time-Momentum Representation.} 
%%%%%%%%%%%%%%%%
%%%%%%%%%%%%%%%%
%%%%%%%%%%%%%%%%
An approach    revived in many recent HVP computations \cite{Francis:2014hoa,Feng2013a,Blum2015} proposes an alternative to the computation of the HVP in momentum space captured in  Eq. \ref{eq:2pt}. 
One starts from the spatially summed vector correlator $G(x_0)$, and writes 
a renormalized HVP as $\hat{\Pi}(Q^2)=\frac{1}{Q^2}\int_{0}^{\infty}dx_0 G(x_0)[Q^2x_0^2-4\sin^2(\frac{1}{2}Qx_0)]$. The challenge is here transferred to understanding the large time behavior of $G(x_0)$. 
It can be shown that the TMR representation is equivalent to the previously discussed Time Moments approach \cite{WittigPoS}.\\
%%%%%%%%%%%%%%%%
%%%%%%%%%%%%%%%%
%%%%%%%%%%%%%%%%

\noindent
{\it Analytic Continuation.} 
%%%%%%%%%%%%%%%%
%%%%%%%%%%%%%%%%
The method discussed in Ref. \cite{Feng2013a}  computes the $\amuhvp$ at small space-like momenta, simultaneously with  the time-like momenta. 
This is achieved by keeping the spatial Fourier transform of  Eq. \ref{eq:2pt} and only integrating in  the 
time direction with a factor $e^{wt}$,
 such that continuous values of $Q^2$, Q=(\vec{q},-iw), are accessed:
$
\bar{\Pi}(Q^2)(\delta_{\mu\nu}Q^2-Q_{\mu}Q_{\nu})=\int dt ~e^{wt}\int d^3\vec{x}~e^{i\vec{q}\vec{x}} ~ \langle J_{\mu}^{em}(\vec{x},t) J_{\nu}^{em} (\vec{0},0) \rangle$. 
$w$ is a  continuous parameter, and $\bar{\Pi}(Q^2)$ is a modified HVP form factor, which has been proven to 
lead to an evaluation of the physical $\amuhvp$. 
The integration range is divided into two parts: the interval $[0,Q_{max}]$, where the analytic continuation %of Eq. \ref{eq:analytic} 
is applied and the integration range $[Q_{max},\infty)$ that is treated in a standard way. 
An advantage of this approach is that the model dependence comes only from the region $[Q_{max},\infty)$. Unfortunately, the method does not seem to lead to an increased precision in the overall calculation of $\amuhvp$.
The described approach is utilized in the computation of $\amuhvp$ for $N_f=2+1+1$ QCD with twisted mass fermions by ETMC \cite{Burger2014}. \\
%%%%%%%%%%%%%%%%
%%%%%%%%%%%%%%%%
%%%%%%%%%%%%%%%%
%%%%%%%%%%%%%%%%

\noindent
{\it Hybrid Method} 
%%%%%%%%%%%%%%%%
%%%%%%%%%%%%%%%%
%Originally 
The approach proposed in Ref. \cite{Golterman2014b} allows for better control of the systematics arising from the decreasing precision of lattice data in the range of  low-$Q^2$. 
The integral that gives $\amuhvp$
is this time divided 
into three non-overlapping regions: $[0,Q^2_{low}]$,$[Q^2_{low},Q^2_{high}]$ and $[Q^2_{high},\infty)$ (cf. 
Fig. \ref{fig:summary}). 
One then applies different fits/moment methods  in the range $[0,Q^2_{low}]$, and performs a simple numerical integration in the region $[Q^2_{low},Q^2_{high}]$. 
Taking into account that roughly 80\% of the contribution to the HVP integral comes from $Q^2<0.2\GeV^2$ domain \cite{Golterman2014b}, this approach helps us give a better estimate of the systematics in the computation of $\amuhvp$ by probing different integration boundaries $Q^2_{low}$ and $Q^2_{high}$. 
In the application of this method by RBC/UKQCD \cite{Blum2016a}, the parametrization in the low-$Q^2$ region is varied, alongside the method used to match the parametrization in this region and the numerical methods used to integrate the mid-$Q^2$ region. This allowed obtaining a reliable estimate of the overall systematic error, which for the strange quark, interestingly,  came out to be smaller than the statistical error. 
\subsection{Summary of current lattice evaluations of the connected HVP}
\begin{figure*}
\begin{minipage}{0.32\textwidth}
\hspace*{-0.05\textwidth}
  \includegraphics[width=1.05\textwidth]{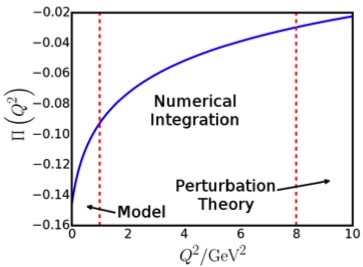}
\end{minipage}
\begin{minipage}{0.32\textwidth}
\vspace*{-4mm}
\hspace*{1mm}
  \includegraphics[width=1.05\textwidth]{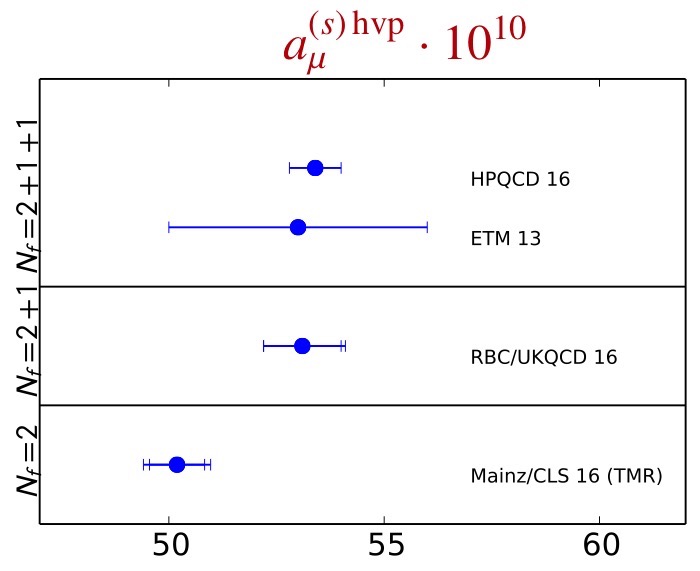}
\end{minipage}
\begin{minipage}{0.32\textwidth}
\vspace*{-3mm}
\hspace*{7mm}
  \includegraphics[width=0.96\textwidth]{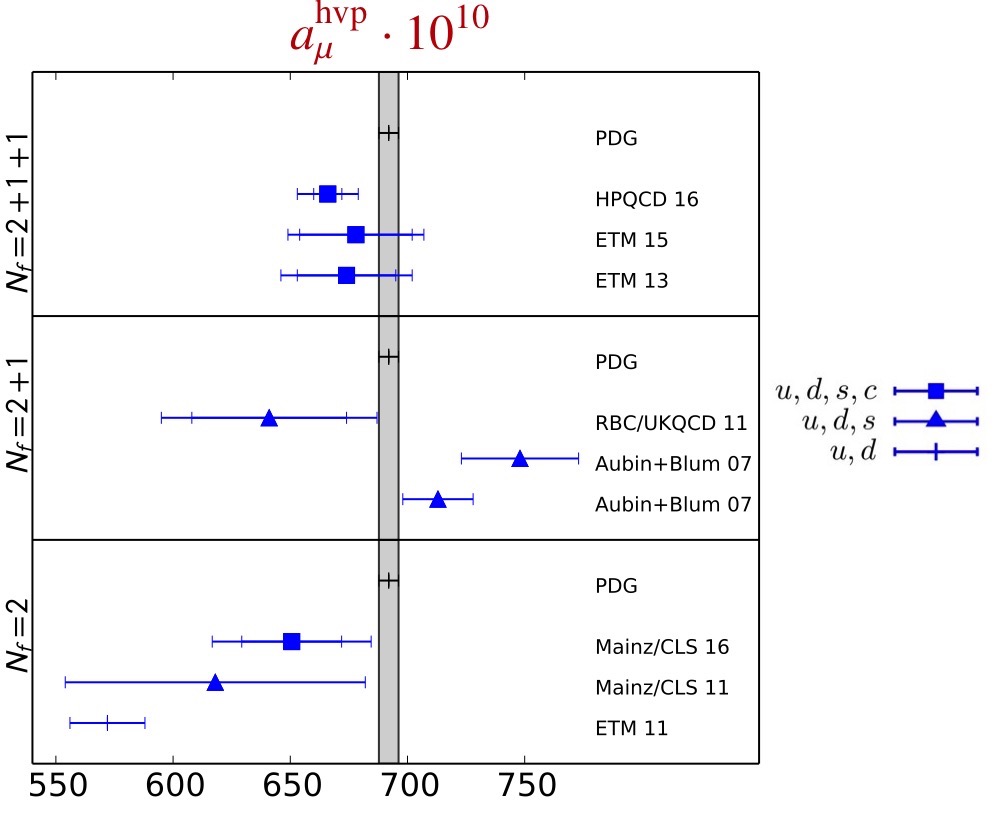}
\end{minipage}
\caption{Left panel:  Illustration of the hybrid  method  \cite{Golterman2014b} reproduced from Ref. \cite{Blum2016a}. Vertical dashed lines represent the integration boundaries: $Q_{low}$, a variable frontier between the parametrization of the low-$Q^2$ region and the intermediate numerical integration region; and the second boundary $Q_{high}$, where the contact with perturbation theory is made. Center and right panels show the summary of strange quark and light quark HVP results by several collaborations using methods described in Sec. \ref{sec:strategies}. The two figures are reproduced from the plenary talk by H. Wittig at Lattice 2016 \cite{WittigPoS}.}
\label{fig:summary}    
\end{figure*}

Fig. \ref{fig:summary} shows the summary of the determinations of different contributions to the HVP by several lattice collaborations previously presented 
 at Lattice 2016 \cite{WittigPoS}. In the left panel the strange contributions by Mainz group on CLS ensembles \cite{HorchPoS}, RBC/UKQCD\cite{Blum2016a}, ETM\cite{Burger2014} and HPQCD \cite{Chakraborty2014} collaborations are compared and  the results  
are mutually consistent. 
The right panel contrasts the PDG estimate of the HVP with the results by various lattice collaborations, which are treating two, three or all four (u,d,s,c) quark flavors dynamically.  
Lattice estimates agree with the PDG value, although the accuracy is still several times lower than the one from non-lattice methods that are the sole contributors to the PDG result. 
%%%%%%%%%%%%%%%%
%%%%%%%%%%%%%%%%
%Summary Figures%%%%%%
%%%%%%%%%%%%%%%%
%%%%%%%%%%%%%%%%
%%%%%%%%%%%%%%%%
%%%%%%%%%%%%%%%%

Recent lattice works on the connected HVP front that are not covered in this review
include the computation of the HVP by introducing a background magnetic field \cite{Bali2015}, a suggestion to perform a subtraction of $\Pi(0)$ using 
the derivative of the twisting angle \cite{deDivitiis:2012vs}, computing HVP directly at an arbitrary
 momentum and in a finite volume (FV) \cite{AntoninProc}, as well as a systematic study of the  FV effects on the BMW ensembles \cite{Malak:2015sla}. Additionally, Ref. \cite{Aubin2016a} gives an estimate of FV effects for the connected HVP from chiral perturbation theory (ChPT) and an interesting proposal to reduce FV effects by twist averaging has been described in Ref. \cite{Lehner2015}. The most recent addition to the rapidly growing literature on lattice determinations of the HVP is the systematic study of the slope and curvature of the HVP by the BMW collaboration \cite{Lellouch2016}, where for the first time continuum limit of these quantities is performed using six different lattice spacings. 
%%%%%%%%%%%%%%%%
%%%%%%%%%%%%%%%%

\vspace*{-2.5mm}
\subsection{Disconnected HVP contribution}
%%%%%%%%%%%%%%%%
%%%%%%%%%%%%%%%%
Quark-disconnected contributions to vector correlation functions (see Fig. \ref{fig:strange}) 
are known to have a bad signal to noise ratio, and represented one of the main sources of unquantified uncertainties in the earlier computations of the HVP from the lattice \cite{Boyle2012,DellaMorte2012}. An estimate of the disconnected HVP from one-loop FV lattice ChPT \cite{DellaMorte:2010aq} and an earlier ETMC work \cite{Feng2011} that included the disconnected contribution for almost half of the ensembles used in the study, 
 have been followed by several contemporary computations of this quantity \cite{Guelpers2014,Chakraborty2016a,Toth2015,Bali2015}. 
The latter have rendered 
this contribution relevant in achieving the aimed sub-percent precision for the total HVP from the lattice. 
Recent work by RBC/UKQCD \cite{Blum2016} is the first to statistically resolve the signal of the disconnected HVP with physical light quarks. The success of the latter is due to pinpointing the mechanisms related solely with the nature of the disconnected contribution, in addition to the application of the noise reduction techniques such as 
all-mode-averaging strategy \cite{Blum2013}, all-to-all quark propagators \cite{Foley2005}, and sparse random sources. Moreover, light-strange quark difference is computed directly, as previously proposed in Ref. \cite{Guelpers2014}, and a result for the disconnected HVP amounting to $\approx1.5\%$ of the total HVP at $3 \sigma$ level is obtained \cite{Blum2016}.
%%%%%%%%%%%%%%%%

\subsection{Isospin breaking correction to the HVP}
\label{sec:iso}
%%%%%%%%%%%%%%%%
%%%%%%%%%%%%%%%%
Similarly to the disconnected piece, the isospin breaking corrections to the HVP are expected to be at a few percent level. However, the goal precision for the HVP from the lattice cannot be achieved before we control the effect of performing the lattice computation in the isosymmetric limit: with up ad down quark masses degenerate and no electromagnetic effects taken into account. Up to now, only a limited number of efforts have been invested in lattice community, in order to determine the isospin breaking effects to the hadronic observables in high precision lattice simulations (for a recent review see \cite{PatellaPoS}). 
The existing estimates of the isospin breaking corrections to the HVP based on phenomenological input \cite{Jegerlehner2011,Chakraborty2016b}, % based on EFT and perturbation theory 
speak in favor of the importance of this correction 
for reaching a  sub-percent precision of the HVP from the lattice. 
In particular, the error budget of \cite{Chakraborty2016b} quotes precisely the unaccounted isospin breaking effects as the largest contribution to the total uncertainty. The definite answer regarding the significance of these effects can be obtained only by carrying out a lattice calculation and several such studies are on the way \cite{MyProc,VeraProc}.
%%%%%%%%%%%%%%%%
%%%%%%%%%%%%%%%%

\subsection{HVP from  new experiments  combined with lattice QCD}
%%%%%%%%%%%%%%%%
%%%%%%%%%%%%%%%%
Note that an alternative way to write Eq. \ref{eq:hvp} features the running of the fine structure constant: $\amuhvp =  (  \frac{\alpha}{\pi} ) \int_0^{1} dx (1-x)
\Delta \alpha_{had}(-Q^2)$ \cite{Lautrup1968}, where $Q^2=\frac{x^2m_\mu^2}{1-x}$. 
Recently, a couple of novel approaches to perform a direct space-like measurement of the
hadronic contributions to the running of the fine structure constant, $\Delta \alpha_{had}(-Q^2)$, 
have been proposed \cite{CarloniCalame2015,Abbiendi2017}. 
A precise value of $\Delta \alpha_{had}(-Q^2)$ could be obtained either in Bhabha scattering with a small modification of the existing experiments \cite{CarloniCalame2015}, or in the recently proposed measurement of the scattering of the high-energy muons on the fixed electron target \cite{Abbiendi2017}. 
The statistical precision of the value of $\amuhvp$ obtained from the latter of the two proposed experiments is estimated to be O(0.3\%) after two years of data taking, while 
the systematics still need to be accounted. 

The original proposals assume using time-like dispersive data and/or perturbation theory beyond the maximal space-like momenta that can be achieved in  space-like scattering experiments, $Q^2_{exp}$.  Similarly to the Hybrid Method described in Sec. \ref{sec:strategies}, one could instead combine the input from the newly proposed experiments 
in some low momentum integration range $[0,Q^2_{exp}]$ 
with lattice QCD calculation of the contribution to the $\amuhvp$ from the region $[Q^2_{exp},Q^2_{high}]$, 
and finally make contact with perturbation theory at some value $Q^2_{high}$.
Due to the complementarity of the integration ranges where the proposed space-like experimental measurements and lattice QCD give precise results, the value of $\amuhvp$ obtained by combining the two methods is expected to be comparable to the projected accuracy of the $g-2$ experiments at Fermilab and J-PARC. 
%%%%%%%%%%%%%%%%
%%%%%%%%%%%%%%%%
%%%%%%%%%%%%%%%%
%%%%%%%%%%%%%%%%

\vspace*{-2.5mm}
\section{Hadronic Light by Light contribution}
%%%%%%%%%%%%%%%%
%%%%%%%%%%%%%%%%
\label{sec:HLbL}
The value for the Hadronic Light by Light contribution entering the phenomenological estimate of $\amu$ \cite{Jegerlehner2009} (cf. Table \ref{tab:amu}) is based on model
%\footnote{This value is also known as the  {\it Glasgow consensus}, as it has been agreed upon at the {\it Topical Workshop on The Muon Magnetic Dipole Moment $(g-2)_{\mu}$} in October 2007. } 
calculations and lattice methods allow for the model independent evaluation of these estimates. 
In the previously described HVP calculations, one would first separate QED and QCD contributions, such that only the QCD piece is computed on the lattice.
 Following a similar approach for HLbL would require to resolve the four point correlation function 
$a_{\mu}^{HLbL} \propto \int dQ^2 f'(Q^2) \sum_{x_1,x_2,x_3,x_4}e^{iQx}\langle J_{\mu}(x_1) J_{\nu}(x_2) J_{\rho}(x_3) J_{\sigma}(x_4) \rangle$ from the lattice, which is computationally very demanding. 
A feasible methodology to compute the HLbL contribution to $\amu$ by factorizing the QED part has been first proposed by T.Blum, T. Izubuchi and M. Hayakawa \cite{Hayakawa2006}. 
An original approach of combining lattice QCD and stochastically generated photon propagators  \cite{Hayakawa2006,Blum2015} has recently been superseded with a more promising approach of  replacing stochastic QED with the analytic photon propagators\cite{Blum2017,Blum2016d}.
On the other hand, recent progress in the dispersive-based analysis inspired by  a data-driven HVP calculations \cite{Colangelo2014a,Colangelo2015a,Colangelo2017c,Pauk2014b}
has motivated lattice practitioners to revisit the mentioned approach of calculating HLbL by computing the four-point correlator on the lattice. Its recent implementation by the Mainz group assumes the computation of the multidimensional integral over a  position-space QED kernel in the continuum \cite{Green2016}. This proposal has recently been tested by evaluating the $\pi_0$-pole contribution to the HLbL scattering amplitude \cite{Green2017} and a more complete calculation is underway. 

In the lattice calculation of the HLbL contribution, again both the connected and the disconnected pieces have to be taken into account.
Despite the tremendous advances the computations of connected HLbL have seen in the previous years, the disconnected contribution still remains a challenge, particularly due to the fact that unlike for the HVP, for the HLbL there is no reason to expect its suppression with respect to the connected part. This concern has been confirmed in  a recent  calculation by RBC \cite{Blum2017}, which estimates the leading disconnected piece for HLbL to be as as much as $\approx50\%$ of the connected one. 
%%%%%%%%%%%%%%%
%%%%%%%%%%%%%%%%
%%%%%%%%%%%%%%%
%%%%%%%%%%%%%%%%

\section{Summary and Outlook}
%%%%%%%%%%%%%%%%
%%%%%%%%%%%%%%%%
Due to its high sensitivity to the potential physics beyond the Standard Model, muon magnetic moment is an optimal channel to unravel new physics. The dominating uncertainties in the theoretical estimate of $\amu$ come from the hadronic contributions: the leading order HVP and HLbL.  
The computation of the connected contribution to the HVP has become a mainstream calculation for lattice QCD by now, and it is particularly encouraging to see that more and more lattice studies are focused on computing the disconnected and isospin breaking corrections to the HVP, as well as the HLbL contributions.
It is expected that once the effect of the isospin breaking to the HVP is known with a precision of O(10\%), the sub-percent accuracy of the HVP contribution from the lattice will already be achieved. 
The experimental precision of the anomalous magnetic moment 
%(currently measured with a precision of 0.54ppm) 
will be improved four times in the next couple of years at the Fermilab new experiment E989 \cite{Venanzoni2016} and at J-PARC, E34 \cite{Saito2012}, making the verification and improvement of the current theoretical prediction of $\amu$ a timely task.
For all these reasons, it is crucial to invest in further development of  the lattice methods described in 
this review and to devise new methods, in order to take full advantage of the first principles calculations of the hadronic contributions to $\amu$. 
\begin{acknowledgements}
I would like to thank the members of the RBC/UKQCD for an enjoyable collaboration on $\amuhvp$ related topics, and to Nazario Tantalo and Fred Jegerlehner for enlightening discussions on the muon g-2. Many thanks to the organizers of Light Cone 2016 for the friendly and stimulating atmosphere of the meeting. Furthermore, I am especially indebted to Massimo Passera and Marija Vranic for valuable comments on the manuscript and to Hartmut Wittig, who generously allowed reproduction of the unpublished figures. 
\end{acknowledgements}
\bibliographystyle{spmpsci}      % mathematics and physical sciences
%\bibliography{refs_lightcone}   % name your BibTeX data base

\end{document}